\documentclass[journal=cmatex,manuscript=article,layout=twocolumn]{achemso}
\setkeys{acs}{maxauthors = 0,email=True}
\usepackage[version=3]{mhchem}
\usepackage[T1]{fontenc}
\usepackage{color,ulem}
\usepackage{amsmath,amssymb}
\usepackage{textcomp}

\usepackage{color,soul} 
\usepackage[english]{babel}
\usepackage[para,online,flushleft]{threeparttable}

\makeatletter
\let\l@addto@macro\relax
\makeatother
\usepackage[fontsize=10pt]{scrextend}
\usepackage{color,ulem}
\usepackage{ulem}
\let\oldmaketitle\maketitle
\let\maketitle\relax
\usepackage{caption}
\usepackage{subcaption}
\usepackage[dvipsnames]{xcolor}
\mciteErrorOnUnknownfalse

\author{Suvo Banik}
\affiliation{Center for Nanoscale Materials, Argonne National Laboratory, Lemont, Illinois 60439, United States}
\alsoaffiliation{Department of Mechanical and Industrial Engineering, University of Illinois, Chicago, Illinois 60607, United States}

\author{Troy D Loeffler}
\affiliation{Center for Nanoscale Materials, Argonne National Laboratory, Lemont, Illinois 60439, United States}
\alsoaffiliation{Department of Mechanical and Industrial Engineering, University of Illinois, Chicago, Illinois 60607, United States}

\author{Rohit Batra}
\affiliation{Center for Nanoscale Materials, Argonne National Laboratory, Lemont, Illinois 60439, United States}

\author{Harpal Singh}
\affiliation{Research and Development, Sentient Science Corporation, West Lafayette, United States}

\author{Mathew Cherukara}
\affiliation[3]{Advanced Photon Source, Argonne National Laboratory, Lemont, Illinois 60439, United States}

\author{Subramanian KRS Sankaranarayanan}
\email{skrssank@uic.edu}
\affiliation{Center for Nanoscale Materials, Argonne National Laboratory, Lemont, Illinois 60439, United States}
\alsoaffiliation{Department of Mechanical and Industrial Engineering, University of Illinois, Chicago, Illinois 60607, United States}

\title{Learning with Delayed Rewards - A case study on inverse defect design in 2D materials}


\begin{document}

\def\pg#1{\textcolor[rgb]{0,0,1}{#1}}
\def\vladan#1{\textcolor[rgb]{0.1,0.6,0.1}{#1}}
\def\sm#1{\textcolor[rgb]{1,0,0}{#1}}

\twocolumn[
\begin{@twocolumnfalse}
\oldmaketitle
\begin{abstract}
Defect dynamics in materials are of central importance to a broad range of technologies from catalysis to energy storage systems to microelectronics. Material functionality depends strongly on the nature and organization of defects – their arrangements often involve intermediate or transient states that present a high barrier for transformation. The lack of knowledge of these intermediate states and the presence of this energy barrier presents a serious challenge for inverse defect design, especially for gradient-based approaches. Here, we present a reinforcement learning (Monte Carlo Tree Search) based on delayed rewards that allow for efficient search of the defect configurational space and allows us to identify optimal defect arrangements in low dimensional materials. Using a representative case of 2D MoS$_2$, we demonstrate that the use of delayed rewards allows us to efficiently sample the defect configurational space and overcome the energy barrier for a wide range of defect concentrations (from 1.5\% to 8\% S vacancies) – the system evolves from an initial randomly distributed S vacancies to one with extended S line defects consistent with previous experimental studies. Detailed analysis in the feature space allows us to identify the optimal pathways for this defect transformation and arrangement. Comparison with other global optimization schemes like genetic algorithms suggests that the MCTS with delayed rewards takes fewer evaluations and arrives at a better quality of the solution. The implications of the various sampled defect configurations on the 2H to 1T phase transitions in MoS$_2$ are discussed. Overall, we introduce a Reinforcement Learning (RL) strategy employing delayed rewards that can accelerate the inverse design of defects in materials for achieving targeted functionality.

\end{abstract}
\end{@twocolumnfalse}
]


\section{Introduction}


Defect dynamics play a significant role in electronic, optical, mechanical, and chemical properties across a wide range of materials\cite{lahiri2010extended,doi:10.1021/acs.jpcc.8b05667}. With proper optimization and design\cite{doi:10.1021/acsomega.6b00500}, these defected structures can yield superior properties. Thus, defect engineering is of significant interest in material design and synthesis\cite{doi:10.1021/acsomega.6b00500,doi:10.1021/acsnano.0c05267,doi:10.1021/acsnano.1c02775,manna2016comparative}. Transition-metal dichalcogenides (TMDs) tend to exhibit exotic properties with a major potential of being applicable in thermoelectrics and catalysis to nanoscale devices\cite{C9NR02873K,fair2015phase,manzeli20172d,doi:10.1021/nl903868w,C8TA04933E,wilson1969transition,mak2016photonics,D1TA00999K}. In 2D transition metal dichalcogenides (TMDs), spatial distribution and dynamics of these defects impact their properties significantly\cite{doi:10.1021/nn500044q,yu_towards_2014,sangwan_multi-terminal_2018,doi:10.1021/acsami.7b02739,fair2015phase,GENG202091}.The most abundant type of defect in TMDs, such as MoS$_2$, is the chalcogen (Sulphur) mono-vacancies\cite{doi:10.1021/acsami.5b01778,doi:10.1063/1.4830036}. During processing or operation of TMD based devices, these point defects are known to transform to lower-energy extended defects, such as line defects\cite{patra_defect_2018,lin_atomic_2014}. 
Such defect mediated transition can result in a cross-over between 2H (semiconductor) and 1T (metallic) phases of MoS$_2$ \cite{patra_defect_2018,lin_atomic_2014,doi:10.1063/1.5040991,PMID:27974834,doi:10.1021/jp2076325,C5CS00151J,doi:10.1063/1.4954257}. From the perspective of emerging applications such as neuromorphic computing\cite{zhang2020perovskite}, it is highly desirable to attain a fundamental understanding of the atomic-scale structure and dynamics of defects in 2D TMDs, as well as their role in driving such structural phase transitions.

Identifying the optimal arrangement of defects and their evolution is a longstanding problem. There are several challenges that need to be addressed in this regard. First, the timescales over which these defects rearrange - these often extend up to several seconds and are clearly not accessible to atomistic simulation techniques. In this respect, structural search algorithms such as genetic algorithms\cite{KIM2021110067,article,LIAO2020414} have been successfully employed to find thermodynamically-favored optimal arrangement of defects in materials \cite{patra_defect_2018}. Second, the transition pathways from point to extended defects are often accompanied by a considerable energy barrier. This prevents search algorithms from efficiently exploring the defect configurational space - the large number of intermediate configurations with increasing barrier precludes the exploration of the optimal defect configuration as shown in Fig 1. Third, there are numerous confounding sub-optimal solutions (`metastable' defect configurations) that are plausible. For example, there are localized defect arrangements (termed as dalmatian effects) as shown in Fig. 1(b) - these local minimas act as sub-optimal traps during the initial optimization stage preventing an exhaustive exploration of defects phase space. Finally, it is worth noting that the variations in defect energetics are often subtle (a few meV/defect). When exploring such defect configurations, it is worth noting that the variance in energy becomes high even with minimal variance in the configurations (see Fig. \ref{fig4}). These complexities in defect energetics and dynamics can significantly delay or result in failure of convergence. There is a clear need to develop and deploy new search algorithms that efficiently navigate the search space and allow convergence to a global minimum with minimal evaluations.


\begin{figure*}[!t]
\centering
\includegraphics[width=\linewidth]{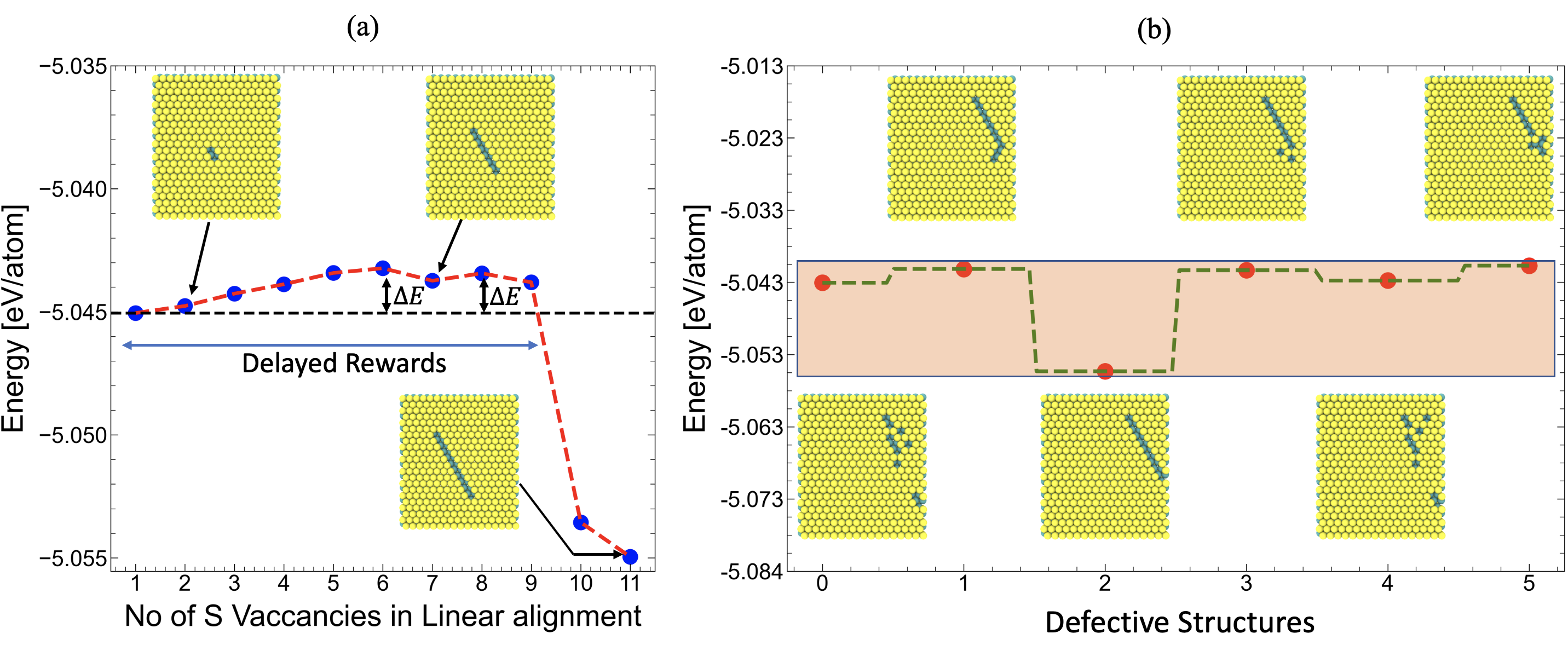}
\caption{(a) Energy barrier encountered during linear alignment of S vacancies. (b) Typical variance in energy against variations in defect configurations of sulphur vacancies (at concentration of 1.5\%) in MoS$_2$.}
\label{fig1}
\end{figure*}

Evolutionary algorithms like Genetic algorithms (GA)\cite{doi:10.1021/acs.macromol.6b01747,mitchell_1998,10.1145/2739482.2764655,LIAO2020414,Goldberg1988GeneticAI,article} are great tools for inverse problems, especially for exploring multidimensional search space and finding crucial structural properties of systems with fair complexity \cite{KIM2021110067,doi:10.1063/1.4897337}. However, a major issue with these algorithms is that they are always driven by the rule of `survival of the fittest’ and tend to favor structures with a better objective in subsequent stages of the search. Although the presence of variation and mutation enables it to explore a wide range of the search space, the convergence is often sluggish, especially when navigating a complex landscape (e.g. defect design) where the path to the absolute global minima in the search presents high enough barrier. In such cases, the search tends to simply divert from the path based on the current state value of the objective which either slows the convergence significantly or results in a failure to reach solution convergence (discussed in detail later).

The number of evaluations, i.e., the time to convergence is an important criterion in materials design. This is especially important since the search engine is often interfaced with a computational model, which is used to evaluate the objective function. Typically, the computational model is either based on density functional theory (DFT) or a classical interatomic potential, which is used to compute the property specified in the objective. Even with the current leadership computing resources, DFT is computationally expensive compared to the classical models and is often limited to smaller system sizes (a few hundred atoms). While classical models are considerably cheaper, it is still important to achieving the target solution in any inverse design problem in as few evaluations as possible.

In this context, reinforcement Learning (RL)\cite{sutton2018reinforcement,mnih_human-level_2015,Popovaeaap7885}has a great potential to facilitate materials design and discovery, and are particularly suitable for this class of problems. Being able to learn on the fly by actively interacting with the environment
makes RL methods highly adaptable, and allow them to make decisions by balancing the exploration-vs-exploitation trade-off.
RL with its ability to explore the state space can thus identify and learn the best behavior of a system based on the past experience gained from interaction with the environment.

Here, we introduce a decision tree-based RL algorithm, i.e., Monte Carlo Tree Search (MCTS) \cite{880a6fbf93da4f0ba5ff61ab3c0aab4e,10.1007/11871842_29,Loeffler_2021,batra2020emerging} which is a powerful machine learning\cite{kocsis2006machine} tool that has found tremendous success in high dimensional (and seemingly intractable) search spaces like in games (such as Chess, Shigo, and Go) \cite{silver2016mastering}, synthesis planning or drug discovery \cite{wang2020towards,Segler2018PlanningCS}, complex materials design and discovery\cite{doi:10.1021/acsomega.9b01480, doi:10.1080/14686996.2017.1344083, D0NR06091G,kajita_kinjo_nishi_2020,gold_ac}. We first demonstrate that the problem of defect design in low dimensional materials is associated with intermediate configurations that pose an energy barrier before reaching the low-energy optimal defect configuration. This material problem is akin to the concept of ``delayed rewards''\cite{watkins1989learning} in RL, which may often take a long sequence of actions, receiving insignificant reinforcement, and then finally arrive at a state with high reinforcement. We interface the MCTS model with a reactive model (ReaxFF)\cite{ostadhossein_reaxff_2017}, which is used to evaluate the energetics for various defective configurations of a representative 2D material i.e. MoS$_2$. Our goal is to start from an initial randomly distributed S point defects (or vacancies) and navigate the search space of various extended defect configurations to identify the lowest energy optimal defect configuration. We highlight the necessary modifications to the MCTS algorithm to efficiently deal with the inverse problems that have ``delayed rewards'' and demonstrate the effectiveness of the approach in defect optimization for a range of different S vacancy concentrations. We compare the efficacy of our approach with our previous work\cite{patra_defect_2018} using GA for defect design. Finally, we provide our perspectives and the applicability of our RL approach for a broad range of inverse problems in materials design and discovery.

\section{Computational Details: MCTS with delayed rewards}
\begin{figure*}[!t]
\centering
\includegraphics[width=\linewidth]{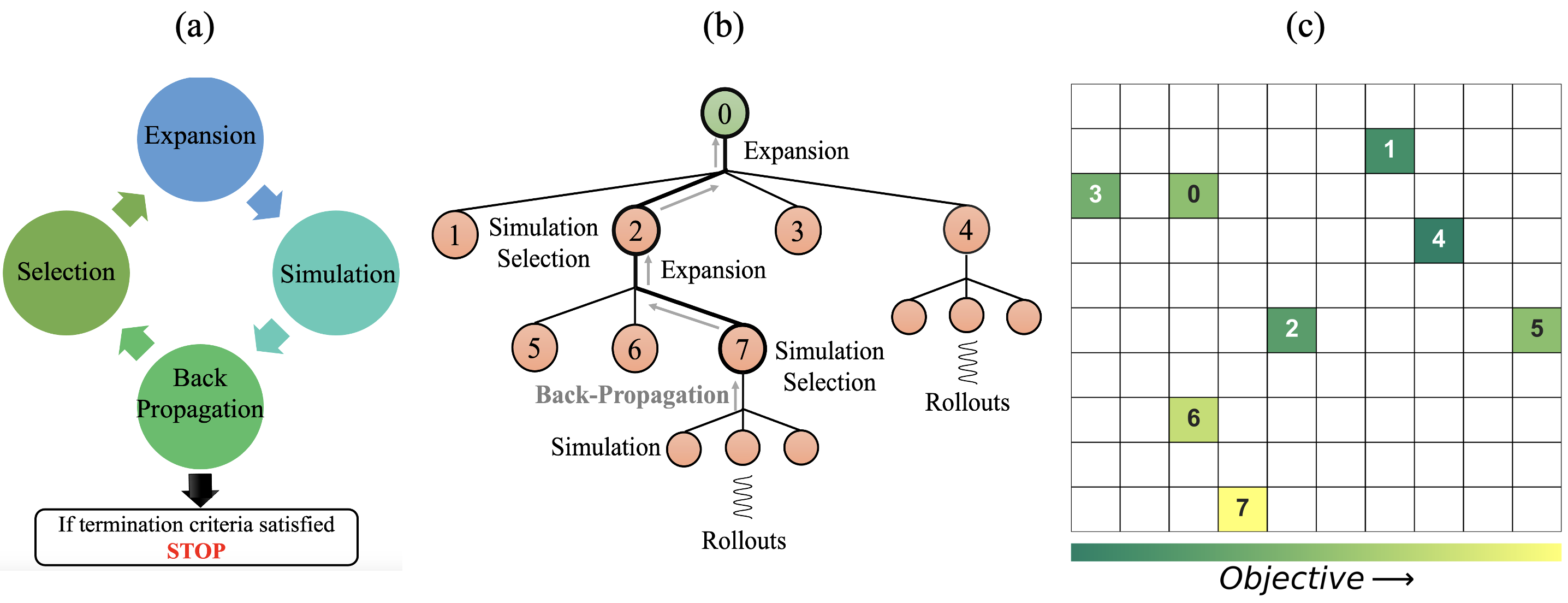}
\caption{(a) The four basic stages of the MCTS algorithm. (b) An example MCTS tree and (c) its corresponding schematic variation in the objective score for each of the node configurations in the discrete action space. The MCTS tree grows (Expansion) and performs simulations (Rollouts) to get a quantitative understanding of the newly added child leaf and learns (Back-propagation). It then selects ideal leaves (Selection) from objective to go deeper in the tree, until the termination criterion is reached. MCTS goes up the energy barrier (e.g., path 0-2-7 in (b), wherein node 7 has higher energy from the rest) to converge to the global minima.}
\label{fig2}
\end{figure*}

The basic MCTS methodology incorporates mainly four stages: `Expansion', `Simulation', `Back-propagation' and `Selection', as shown in Fig. \ref{fig2}. The `Expansion' stage is where the tree is grown by adding child nodes that correspond to perturbations of parent parameters. In the next Simulation stage, a finite number of rollouts are carried out at the newly created leaf node based on a predefined objective. In the `Backpropagation' stage the current node and its all predecessors are updated with rollouts information. This is to get a qualitative objective of the decedent leaf nodes. The `Selection' phase is driven by a popular tree policy upper confidence bound for parameters (UCB)\cite{10.1007/11871842_29}. The UCB of a leaf node is defined as

\begin{equation}
{\rm UCB}({\rm node}_{i}) = -{\rm min}(z_{1},z_{2},z_{3},....z_{i})+C\sqrt{\frac{\ln(v_{p})}{v_{i}}}
\label{eq1}
\end{equation}

\noindent where $z_{i}$ and $v_{i}$ are the reward and the visit count of the $i^{th}$ node, respectively. $v_{p}$ is the visit count of the parent node and $C$ is a constant to balance exploration and exploitation. The value of $C$ can be controlled adaptively based on the progress of the search. The reward was set to the energy of the minimized configurations in meV.

\begin{figure*}[!t]
\centering
\includegraphics[width=\linewidth]{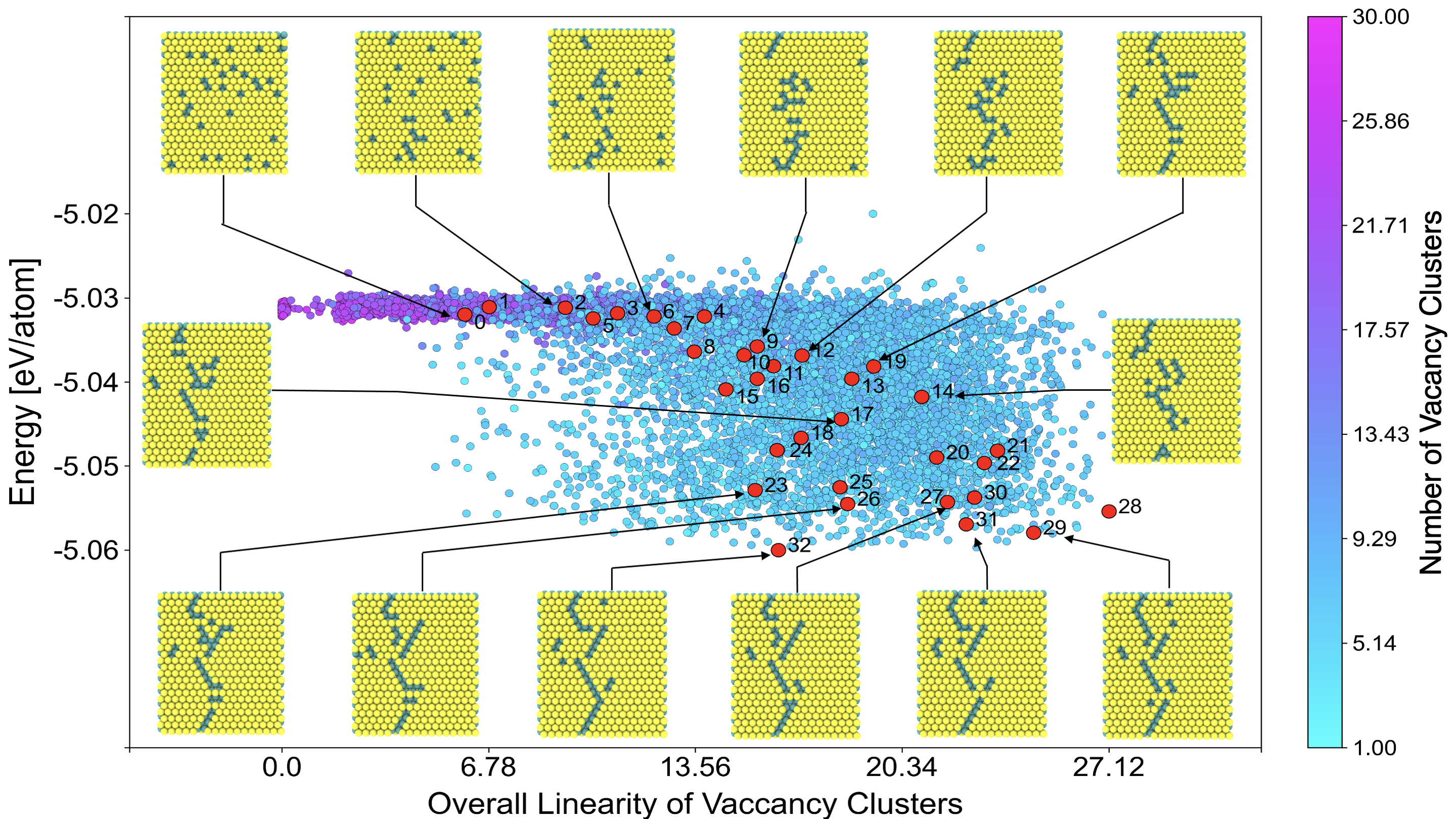}
\caption{MCTS trajectory from the root node to the terminal node showcasing the evolution of the configurations for the case of 4\% S vacancy concentration. Similar plots for other vacancy concentrations of 1.5\%, 5\%, and 7.5\% are provided in Supplementary Information Fig. S2, S4, and S6.}
\label{fig3}
\end{figure*}

We employed the MCTS as an AI optimizer to search for the energetically most favorable alignments of point defects starting from a random initial defect of S vacancies distributed in the chalcogen layer of the MoS$_2$ system (see Fig.\ref{fig3}). As stated earlier, the inherent challenge with MoS$_2$ system is that it has a considerable initial energy barrier for linear alignment of vacancies, although this event is energetically favorable; see Fig. \ref{fig1}(a). Furthermore, the formation of the line defect in a large unit cell is statistically unlikely. There are many local minimas that are both and statistically more likely to form during a search than the more stable linear configurations; Fig. \ref{fig1}(b). We find these local minimas to be an impediment for the GA-based algorithms which rely on rare random events during the course of the optimization in order to form a sufficiently large line defect and allow such structure to dominate the GA's generational population pool.

To address this issue, we modify the UCB function to incorporate the concept of delayed rewards in this optimization problem. A delayed reward to the objective helps the optimizer to initially explore the search space thoroughly to sample enough configurations which are energetically as well as configurationally favorable of forming line defects during the successive stages of the search algorithm and thus converging to the global minima of the search space. We augment the $C$ parameter in UCB equation
with a term that scales with respect to the structural uniqueness of the node and total overall energy evaluations till the specific point of the search considered. This function, $g(C,N)$, is given by the relationship,

\begin{equation}
g(C,N)= 
\begin{cases}
    C*e^{-(\alpha N)^2},& \text{if } E_{N}<E_{(N-1)}^{\rm best(K)}\\
    C*e^{-(\alpha K)^2},              & \text{otherwise}
\end{cases}
\label{gcn}
\end{equation}

\noindent where $C$ is the same exploration constant as included in Eq. \ref{eq1} (and specified by the user). This serves as the initial value of $g(C,N)$ prior to the rewards are being discounted. This also serves as the upper bound of this function. $N$ is the total number of energy evaluations performed by the MCTS algorithm and $\alpha$ is a user-defined constant. $E_{(N-1)}^{\rm best(K)}$ is the lowest energy configuration, found at the $K^{th}$ MCTS evaluation, up until the total of $N-1$ total energy evaluations, while $E_{N}$ is the energy of the configuration for the $N^{th}$ evaluation. 
The function $g(C,N)$ tunes the exploration part of the UCB (Eq.\ref{eq1}) in such a way that during the initial stages of the search the exploration part of the UCB equation dominates the objective score. The $g(C,N)$ term cannot go below a certain value given by Eq. \ref{gcn} for a given $N$ and thus directing the algorithm to initially ignore the immediate reward found during initialization (N is very less) and begin building up a large knowledge base of structures from which it can begin making decisions off. At later stages of the search, the exploration term will quickly decay towards a minimal value upon encountering a highly rewarding configuration. As more is known about the total configurational space the node selections will begin to become biased towards the branches of the Tree which are proving to be highly rewarding. The overall MCTS objective function thus takes the form

\begin{equation}
{\rm UCB}({\rm node}_{i},N) = -{\rm min}(z_{1},z_{2},z_{3},....z_{i})+g(C,N) \sqrt{\frac{\ln(v_{p})}{v_{i}}}.
\label{eq2}
\end{equation}



\section{Results and Discussions}
We start with randomly distributed Sulphur vacancies of concentrations ($\rho$) 1.5\%, 4$\%$, 5$\%$ and 7.5$\%$ on a monolayer of periodic MoS$_2$ system, as shown in Fig. \ref{fig3}.
For each concentration, the atomic interactions in the defective MoS$_2$ structures are modeled using  reactive force field ReaxFF \cite{ostadhossein_reaxff_2017} as implemented in the open-source LAMMPS \cite{plimpton_fast_1995} package. Periodic boundary conditions are employed in the plane of the MoS$_2$ sheet. For evaluations of the MCTS rewards ($z_i$), it is interfaced with LAMMPS. Proposed defect configurations from the MCTS run are minimised using LAMMPS simulations and the obtained energies of the relaxed configurations are passed back to the MCTS run as rewards. The LAMMPS script used for this purpose is included with supplementary information.
We note that in a single layer MoS$_2$, a plane of Mo atoms is sandwiched between two planes of S atoms and in each S plane, the atoms are organized in a 2D triangular lattice. As the inter-layer vacancy migration is associated with a very high energy barrier (> 5 eV) and also previous DFT calculations\cite{doi:10.1021/nl4007479} report that the formation energies of S-vacancy (2.12 eV) are lower than all other types of defects including anti-sites, we restrict our search space to the top layer of S atoms. The defect density is defined as the ratio of vacant S sites to the total S cites in the top layer of a MoS$_2$ film. Here, our objective is to obtain the configuration with the lowest energy, i.e., linear alignment of S vacancies. Fig. \ref{fig3} demonstrates the evolution of the configurations that are discovered during MCTS for the case of 4\% vacancy concentration, wherein configurations from root node (depth 0) to the terminal node (depth 32) are shown. While the starting configurations at lower tree-depth values are seen to be randomly arranged vacancies, the final configuration, with line defects, is obtained at higher tree-depth value of 32.

During the search, it was observed that most of the defects tend to come together to form aggregates (Fig. \ref{fig3}) which are defined as vacancy clusters in this study. These clusters may have vacancies aligned as line defects. In Fig. \ref{fig1}(a), we can see that, although there is an initial energy barrier associated with these line defects, once a sufficient number of vacancies are in line the overall energy sharply drops with increment in the size of the line defects. It also to be noted that this feature of forming a line makes the initial small linear aggregates of special interest (even though it is of high energy) because of their potential of yielding an energetically low offspring upon further perturbation as a parent. This makes the `Overall linearity' of vacancy aggregates as a metric that can provide either the idea of the potential of a configuration of becoming a good parent or the configuration being energetically low itself. Thus, two metrics were used to quantify the different levels of vacancy alignment in the configurational space. The first is the overall linearity of the vacancy clusters, which is given by the expression:


\begin{equation}
{\rm Overall~linearity}  =  \sum_{i=1}^{N_{c}} n_{i} \ell_{i}   
\label{eq3}
\end{equation}

\noindent where $N_{c}$ is the total number of isolated vacancy clusters, $n_{i}$ is the number of S vacancies in the $i^{th}$ cluster and $\ell_{i}$ is the linearity of $i^{th}$ vacancy cluster which can have a value between 0-1 based on how linearly the vacancies are aligned. For our case, a value of 0 was assigned to the clusters where $\ell_{i}$<0.9. The second metric we employed is the total number of isolated vacancy alignments present in a configuration. The lesser the number of isolated vacancies, the higher is the vacancy alignment - such configurations tend to be energetically lower.

As captured in Fig. \ref{fig3}, the search starts at node depth 0 (head node) with a configuration having random vacancies alignment and keeps on going deeper into the tree via expansion and selection till node depth 32, where the most energetically favorable configuration is obtained. We introduce 4 different kinds of moves to create an offspring leaf node by perturbing its parent node. These are swap, shift, associate, and dissociate. The `swap' moves randomly perturb the vacancies and allow local alignments to form, the `shift' move perturbs randomly selected vacancies within their local neighborhood, `associate' move brings the isolated vacancies to align with the other vacancy arrangements formed and the `dissociate' move breaks an alignment by moving randomly selected vacancies from the alignment. These moves are shown pictorially in Supplementary Information Fig. S1. During the MCTS run, a move is selected based on the specific probabilities and applied to the parent node to generate offsprings for playouts or expansion. The choice of probabilities for these different moves are based on the concept of delayed rewards. A detailed description of the probabilities of selection associated with these moves is provided in the supplementary information.
A depth-based scaling was used to scale down the number of vacancies perturbed with the increasing depth of the tree. We can clearly see from Fig. \ref{fig3} that, as the depth of the tree increases, the configurations tend to become more alike in terms of vacancy arrangement as the energy is decreasing. Perturbing more vacancies will introduce more randomness in the offspring leaf node configurations which might lead to a delay in convergence.


\begin{figure*}[!t]
\centering
\includegraphics[width=\linewidth]{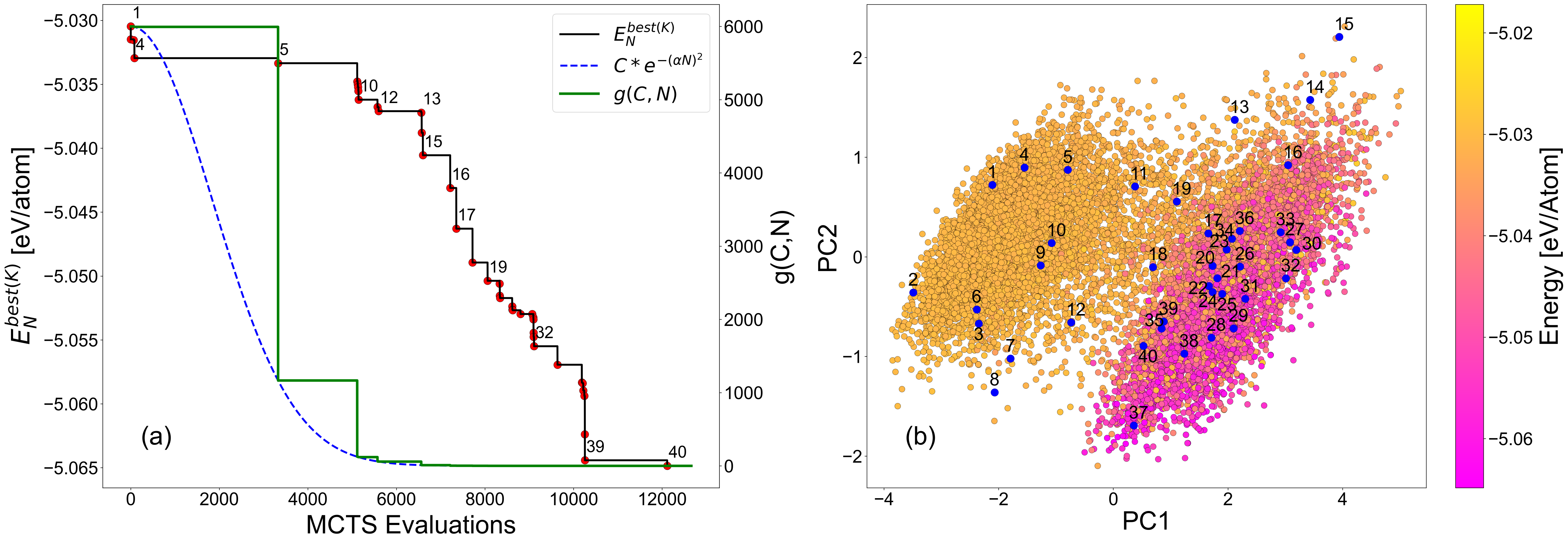}
\caption{Evolution of the (a) energy of the best candidate $E^{\rm best(K)}_{N}$ and the $g(C,N)$ of the exploration part in Eq. \ref{eq3} as a function of the MCTS evaluations for the case of 4\% vacancy concentration. (b) Representation of all the MCTS configurations on the principal component (PC) space derived using the SOAP fingerprinting scheme for the case of 4\% vacancy concentration. Similar plots for other vacancy concentrations of 1.5\%, 5\%, and 7.5\% are provided in Supplementary Information Fig. S3, Fig. S5 and Fig. S7.}
\label{fig4}
\end{figure*}

The probability of selection of the mentioned moves in accordance with the tree depth also plays a crucial role in the convergence of the MCTS search. A more chaotic perturbation move like swap, if used very deep inside the tree while the search is approaching convergence, will introduce unnecessary randomness in the offspring configurations. Likewise, the frequent use of local perturbation move like shift, associate, etc. in a very shallow tree will cause a loss of diversity in the initially sampled configurations. In the long run that might prove detrimental to the search convergence. Hence we resorted to two schemes for getting better performance from the MCTS search (a) depth-based perturbation scaling (b) appropriate selection of the moves with tree depth.


From Fig. \ref{fig3}, it can be observed that the energy of the configurations does not consistently go down with an increment in the tree-depth (for e.g. 11-12, 18-19 etc.). As shown in Fig. \ref{fig1}, the search needs to climb up the energy barrier to converge to an energetically lower configuration. This trait is extremely crucial for systems like MoS$_2$ where the configurations at a later stage become almost structurally indistinguishable and display only slight differences in vacancy alignment, but having a considerable variance in energy. For example, configurations at depth 27 and 32 in Fig. \ref{fig3} have almost identical vacancy alignment but an energy difference of $\sim6.81$ meV/atom, which is quite high considering the overall energy range of the sampled defect configurations. Decisions solely based on the current state of the search will likely lead to either sluggish or no convergence to the optimal solution.

Next, we observe the effect of delayed rewards on the overall performance and convergence of the search as a whole. Fig. \ref{fig4}(a) shows the energy evolution $E^{\rm best(K)}_{N}$ of the best candidate with MCTS evaluations, the augmented function $g(C,N)$ applied to the objective with respect to the number of energy evaluations. These results are shown for a vacancy concentration of 4\%, while those for 1.5\%, 5\%, and 7.5\% are included in Supplementary Information (Figs. S3, S5, and S7). For the energy ranges considered ($-5100$ to $-5000$ meV/atom), the value of exploration constant $C$ in Eq. \ref{eq2} is set to 6000. The criteria of delayed reward is set in such a way (in Eq. \ref{gcn}) that there is a reduction in $g(C,N)$ term only if a drop in the energy value of the configuration is encountered. To follow the configurational evolution during the MCTS search, we analyzed the structural features of the configurations in fingerprint space. Fig. \ref{fig4}(b) shows the representation of all the MCTS sampled configurations in the first two principal components (PC), although the complete high-dimensional fingerprint space consisted of 147 dimensions.
A SOAP (Smooth overlap atomic positions) \cite{de_comparing_2016,PhysRevB.87.184115} fingerprint from python library DScribe \cite{noauthor_dscribe:_2020} was used for the fingerprint computations. It is evident from Fig. \ref{fig4}(b) that the whole configurational space is divided into two distinct regions. The region on the left is dominated by high energy configurations (roughly $-5.02$ to $-5.035$ eV/atom) while the second region to the right consists of mostly low-energy ($\sim-5.04$ to $-5.065$ eV/atom) configurations, although they still have high variance in energy values. There is a clear configurational gap between these two regions. This corresponds to the energy barrier associated with the formation of an initial linear alignment as seen in Fig. \ref{fig1} that is needed to be overcome before we can arrive at optimal defect configuration.


\begin{figure*}[!t]
\centering
\includegraphics[width=\linewidth]{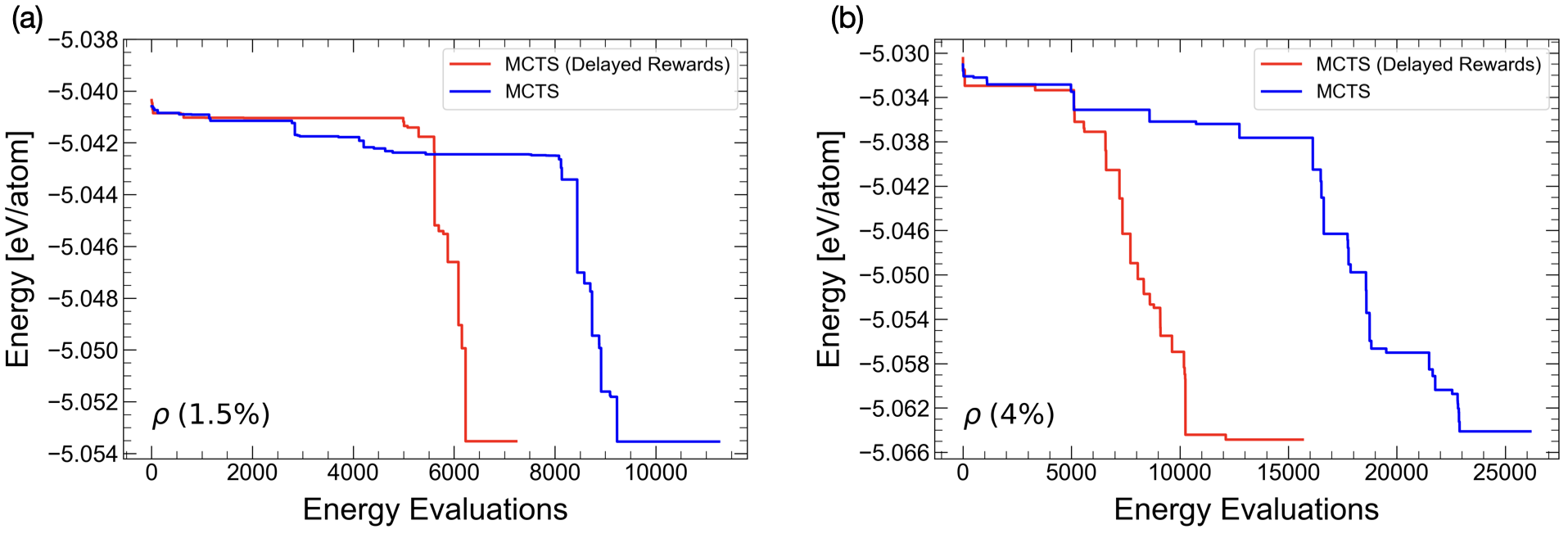}
\caption{Comparison of evolution of the best candidate with number of energy evaluation for MCTS with delayed rewards and without delayed rewards for  S vacancy concentrations ($\rho$) of (a) $1.5\%$, (b) $4\%$.}
\label{fig6}
\end{figure*}

From Fig. \ref{fig4}(a) it can be seen that until about 5000 MCTS evaluations, $g(C,N)$ function isn’t at its minimum and the overall energy of the current best configuration is quite high (-5.035 eV/atom). During this initial exploration phase of the search, the MCTS mainly explores the left region in Fig. \ref{fig4}(b) (points 1-13 in Fig. \ref{fig4}(a)(b)) and samples configurations randomly till a suitable initial alignment of the vacancies is formed. This helps the MCTS search to overcome the initial configurational barrier and move to the zone on the right with lower energy configurations (point 14 or higher in Fig. \ref{fig4}(b) ). At this point, the reward increases and becomes maximum as the search moves to its exploitation phase. During this exploitation phase, the value of the exploration part decreases, and the UCB objective function (Eq. \ref{eq2}) becomes biased towards the rewards (i.e., the energy of the configurations). The energetically lower configurations are exploited frequently to generate new offsprings. After the search moves to the right region in the configurational space in Fig. \ref{fig4})(b), the MCTS performs extremely well (in lowering the objective) due to the fact that it does not only exploits the configurations with the lowest energies but also the configurations with similar vacancy arrangement while having comparatively higher energy. This is because of the subtle balance maintained between the exploration and exploitation by UCB Eq. \ref{eq2}. This helps the search to converge to the possible global minima very quickly even if the optimization problem involves a relatively high energy barrier to the optimal solution.


\begin{figure*}[!t]
\centering
\includegraphics[width=\linewidth]{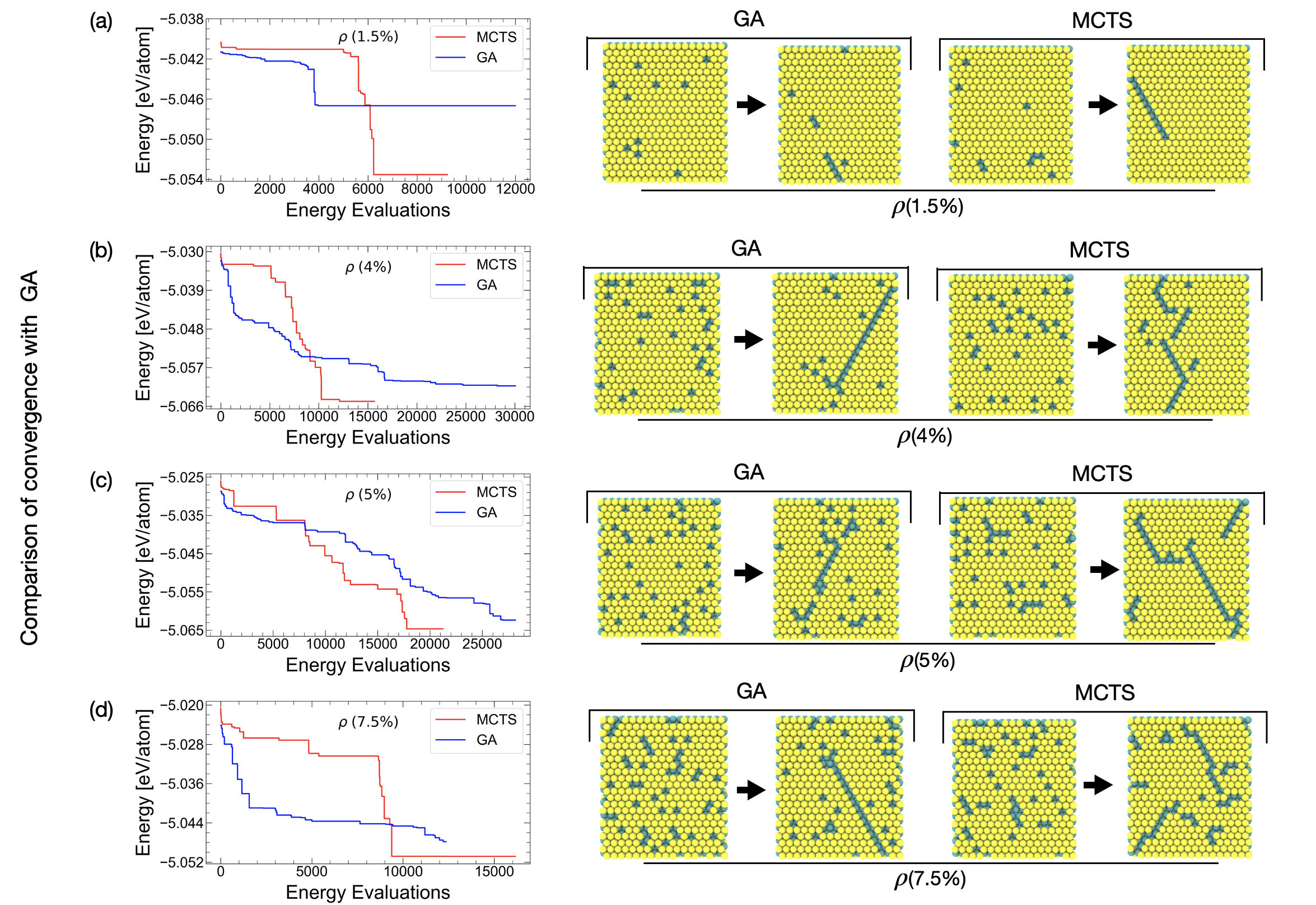}
\caption{Comparison of evolution of the best candidate with number of energy evaluations, along with the initial and final optimised configuration for MCTS (Monte Carlo Tree Search) and GA (Genetic  algorithm) \cite{patra_defect_2018} for 4 different S vacancy concentrations ($\rho$) of (a) $1.5\%$, (b) $4\%$, (c) $5\%$ and (d) $7.5\%$.}
\label{fig5}
\end{figure*}


\begin{table*}[th]
\centering
\setlength{\tabcolsep}{0.8em}
\renewcommand{\arraystretch}{1.2}
\caption{Comparison of MCTS and GA for the search of low energy defect configurations in MoS$_2$. The results of GA are taken from the past work \cite{patra_defect_2018}.}
\label{table1}
\begin{tabular}{c|c|c|c|c}
\hline\hline
 &  \multicolumn{2}{c|}{GA} & \multicolumn{2}{c}{MCTS}\\

\cline{2-5}

$\rho$ & No. of Evaluations   & Energy of the lowest  & No. of Evaluations   & Energy of the lowest  \\
  & to converge  &  configuration [eV/atom]   & to converge  & configuration [eV/atom]\\

\hline
1.5\%   & 3960 & -5.04666  & 6228  & -5.05352\\
4\%   &  28200 & -5.06125  & 12115  & -5.06485\\
5\%   &  26960 & -5.06238  & 17929  & -5.06465\\
7.5\%  & 12200 & -5.04783  & 9378  & -5.05081\\

\hline\hline
\end{tabular}
\end{table*}

We then compare the performance of MCTS with delayed rewards to a regular MCTS search i.e. without any delayed rewards, for two vacancy concentrations of 1.5\% and 4\% (Fig. \ref{fig6}). For the case of 1.5\% vacancy concentration, both the MCTS, with and without delayed rewards seem to perform well while converging to the global minima. For MCTS with delayed rewards, it can be seen in Fig. \ref{fig6}(a) that, initially around $\sim$6000 evaluations the energy of the sampled configurations are comparatively high compared to the MCTS without delayed rewards. In this phase, MCTS is exploring the search space since there is a penalty to the rewards. Once it crosses that stage the energy sharply falls and the search converges with minimal evaluations. However, as the vacancy concentration increases to 4\%, regular MCTS performs sluggishly and takes a large number of evaluations to converge, and the converged configuration is also slightly higher in energy ($\sim$1 meV) than its delayed reward case (see Fig. \ref{fig6}(b)). This is due to the fact that initial inefficient sampling of configuration space, at a later stage, causes a lack of suitable parent configurations from which a configurationally and energetically ideal offspring can be obtained. In this case, MCTS with delayed rewards outperforms MCTS and converges to an energetically lower configuration with significantly fewer evaluations ($\sim$ 12000). It is also to be noted that changing the hyperparameters may result in a variance in the number of evaluations to converge. However, as the vacancy concentration is increasing, MCTS with delayed rewards performs extremely well both in terms of taking fewer evaluations to converge and obtaining energetically lower configurations upon convergence.

Finally, we compare the search performance of MCTS (with delayed rewards) with that of GA \cite{patra_defect_2018} in Fig. \ref{fig5} and Table \ref{table1}. For the case with 4\% vacancy concentration (Fig. \ref{fig5}(a)), the energy of the best configuration from the MCTS search till $\sim$9000 evaluations is comparatively higher than the GA. Till $\sim$5000 evaluations, the MCTS is in its exploration phase but afterward, it moves into the exploitation phase, the overall best energy of the search $E^{\rm best(K)}_{N}$ drops very sharply. By $\sim$9000 evaluation it becomes lower than that of GA. While it takes GA around $\sim$30000 evaluations to converge, we find that the MCTS converges within $\sim$12000 evaluations and successfully finds a configuration that is $\sim$4 meV/atom lower than the best candidate of the GA (Table \ref{table1}). Comparing the best configurations from the GA and MCTS in Fig. \ref{fig5}(a), we can clearly see that the vacancies tend to form extended line defects in both cases. However, compared to GA the configuration obtained from MCTS has almost all the vacancies aligned.

For the case of 1.5\%, 5\%, and 7.5\% vacancy concentrations as well, the final configurations obtained from MCTS, which have almost all the vacancies aligned, are energetically lower compared to those obtained from GA (see Fig. \ref{fig5}). The final configurations for $\rho$ 1.5\% had all of its vacancies aligned as extended line defects while the cases with 5\% and 7.5\% concentration obtained from MCTS seem to have most of the vacancies aligned as line defects unlike those for GA, as captured in Fig. \ref{fig5} (a), (b), and (c). Because of this, MCTS identified configurations are energetically lower than the ones obtained from GA (Table. \ref{table1}). The total energy evaluations taken by MCTS to converge to these configurations are also significantly less compared to that for GA. In Fig. \ref{fig5}(a) for ($\rho$)=1.5\%, it can be seen that GA search is getting saturated after $\sim$ 4000 evaluations, and the energy of the final configuration is considerably high when compared to that of the final configurations from MCTS($\sim$ 6.7 meV). For the case of both vacancy concentrations ($\rho$) of 5\% and 7.5\% (Fig. \ref{fig5}(b) and (c)), the overall energy of the best configuration from MCTS is higher for initial $\sim$9000 evaluations as compared to GA since MCTS is in its exploration phase. However, afterward, as MCTS moves towards the exploitation phase the best candidate energy goes down very sharply. Overall, it takes substantially fewer evaluations for MCTS to converge as compared to that for GA.

The lowest energy configurations for each of the four vacancy concentrations of 1.5\%, 4\%, 5\%, and 7.5\% are shown in Fig. 5 and can be seen to display most of the vacancy defects aligned as a line. During the search, the vacancies tend to form low energy aggregates (Fig. \ref{fig1}(b)). Many of these aggregates might act as local minima and can trap the optimizer algorithm. However, it is also noticeable (in Fig. \ref{fig3}) that some of these vacancy aggregates acts as a precursor to a low energy configuration with line defects. Thus there is clearly a relationship between the occurrence of these vacancy aggregates and the formation of line defects in the subsequent stages. From our search, the final configurations obtained had most of the vacancies aligned as line defects and were energetically lower than those obtained from the competing GA.

The nature of defect aggregation has significant implications for phase transitions in 2D TMCs\cite{patra_defect_2018,lin_atomic_2014,doi:10.1063/1.5040991,chen2016pressure,PMID:27974834,doi:10.1021/jp2076325,C5CS00151J,doi:10.1063/1.4954257}. During the early part of the search, we note that the single vacancies can cluster to form many small aggregates like dimers or trimers - these configurations are energetically higher (i.e. metastable) than those with extended line vacancies but have been found in the experiments \cite{PhysRevLett.102.195505, doi:10.1021/nl9011497}. As shown in our earlier works, these do not trigger the 2H-1T phase transition in MoS$_2$ system with defects\cite{patra_defect_2018,doi:10.1063/1.4954257,lin_atomic_2014}. The extended line defects, which represent energetically lower configurations as identified by our MCTS search, lead to the formation of an intermediate $\alpha$ phase near the defective region. The induced stress causes S atoms to hop towards the defective region. Although there is an energy barrier to be surmounted, the 1T phase is more likely to nucleate near this region. It is also very likely that coupling of two $\alpha$ phase regions at 60$^\circ$ may trigger the formation of 1T domains in 2H phase of MoS$_2$ \cite{lin_atomic_2014}. The presence of these extended line defects tends to aid the transformations from 2H to 1T.

\section{Conclusion}

In summary, we have introduced the concept of using reinforcement learning (RL) algorithms such as MCTS with delayed rewards to accelerate materials search and discovery problems where there either exist a number of unstable intermediates along the search pathway or involve surmounting high energetic barrier to reach optimal configurations. Using a representative and well-studied problem of defect optimization in 2D TMC such as MoS$_2$, we demonstrate that the use of delayed rewards facilitates enhanced exploration as well as the exploitation of the search pathways leading to the identification of optimal defect configurations. For a range of different vacancy concentrations studied, our RL algorithm suggests that the initial randomly distributed S vacancies tend to aggregate and form energetically favorable line defects – the vacancy aggregation process involves an energy barrier of $\sim$3-5 meV/atom that depends strongly on the number of linear S vacancies. We show that the presence of this energy barrier as well as subtle energetics between various low energy defective (and degenerate) vacancy clusters necessitates the use of delayed rewards. The various different MCTS search pathways are analyzed in the fingerprint space to demonstrate the effectiveness of ``learning with delayed rewards''. The various favorable pathways for S vacancy aggregation from an initial randomly distributed point vacancy to an optimal line effect are discussed in detail. We further compare the performance of our MCTS search with that of genetic algorithm (GA) – the MCTS is able to predict lower energy configurations in fewer search evaluations compared to GA. Thus, the speed of the search as well as the quality of the solution obtained is superior for the representative cases considered in this study. Overall, this study provides useful insights into pathways for defect aggregation in low dimensional materials and introduces a search strategy that allows for materials discovery in problems where the search pathways have unstable intermediates or high barrier to the solution.

\section{Supporting Information}

The representative figure of four basic MCTS moves used, Description of selection probability of moves, MCTS trajectory from the root node to the terminal node (for the vacancy concentrations of $1.5\%$, $ 5\%$, and $7.5\%$), the evolution of the energy of the best candidate $E^{best(K)}_{N}$ and the $g(C,N)$ (for vacancy concentrations of $1.5\%$, $ 5\%$, and $7.5\%$), representation of all the MCTS configurations on the principal component space (for vacancy concentrations of $1.5\%$, $ 5\%$, and $7.5\%$). The LAMMPS script used for the minimization of the configurations.


\section{Acknowledgements}

This material is based upon work supported by the U.S. Department of Energy, Office of Science, Basic Energy Sciences- Artificial Intelligence and Machine Learning at DOE Scientific User Facilities program under Award Number 34532. Use of the Center for Nanoscale Materials, an Office of Science user facility, was supported by the U. S. Department of Energy, Office of Science, Office of Basic Energy Sciences, under Contract No. DE-AC02-06CH11357. This research also used resources of the Argonne Leadership Computing Facility at Argonne National Laboratory, which is supported by the Office of Science of the U.S. Department of Energy under contract DE-AC02-06CH11357. This research used resources of the National Energy Research Scientific Computing Center, a DOE Office of Science User Facility supported by the Office of Science of the U.S. Department of Energy under Contract No. DE-AC02-05CH11231. We gratefully acknowledge the computing resources provided on Fusion and Blues, high-performance computing clusters operated by the Laboratory Computing Resource Center (LCRC) at Argonne National Laboratory.


\nocite{*}
\normalem 
\bibliography{ref}

\end{document}